# Unravelling local spin polarization of Zhang-Rice singlet in lightly hole-doped cuprates using high-energy optical conductivity

Iman Santoso[1,2, 12], Wei Ku[3,*], Tomonori Shirakawa[4,5,6], Gerd Neuber[7], Xinmao Yin[1,2], M. Enoki[8], Masaki Fujita[8], Ruixing Liang[9], T. Venkatesan[1,10], George A. Sawatzky[9], Aleksei Kotlov[11], Seiji Yunoki[1,4,5,6, †], Michael Rübhausen[1,7], Andrivo Rusydi[1,2,‡]

[1]*National University of Singapore and Nanotechnology Initiative (NUSSNI)-Nanocore, Department of Physics, National University of Singapore, Singapore 117576*

[2]*Singapore Synchrotron Light Source, National University of Singapore, Singapore 117603*

[3]*Department of Physics and Astronomy, Shanghai Jiao Tong University, Shanghai 200240, China.*

[4]*Computational Condensed Matter Physics Laboratory, RIKEN, Wako, Saitama 351-0198, Japan*

[5]*Computational Quantum Matter Research Team, RIKEN Center for Emergent Matter Science (CEMS), Saitama 332-0012, Japan*

[6]*Computational Materials Science Research Team, RIKEN Advanced Institute for Computational Science (AICS), Kobe, Hyogo 650-0047, Japan*

[7]*Institut für Nanostruktur und Festkörperforschung, University of Hamburg, Luruper Chausse 149, Center for Free Electron Laser Science (CFEL), D-22607 Hamburg, Germany*

[8]*Institute for Material Research, Tohoku University, Katahira, Sendai 980-8577, Japan*

[9]*Department of Physics and Astronomy, University of British Columbia, Vancouver, British Columbia, Canada V6T-1Z1*

[10]*Department of Electrical and Computer Engineering, National University of Singapore, Singapore 117576, Singapore*

[11]*Photon Science Division, Deutsches Elektronensynchrotron (DESY), Notkestrasse 85, 22607 Hamburg, Germany*

[12]*Departemen Fisika Fakultas Matematika dan Ilmu Pengetahuan Alam (FMIPA) Universitas Gadjah Mada, Sekip Utara Bulaksumur BLS 21, Yogyakarta 55281, Indonesia*

*Correspondence to: *weiku@mailaps.org; †yunoki@riken.jp; ‡phyandri@nus.edu.sg;

**Abstract:** Unrevealing local magnetic and electronic correlations in the vicinity of charge carriers is crucial in order to understand rich physical properties in correlated electron systems. Here, using high-energy optical conductivity (up to 35 eV) as a function of temperature and polarization, we observe a surprisingly strong spin polarization of the local spin singlet with enhanced ferromagnetic correlations between Cu spins near the doped holes in lightly hole-doped $La_{1.95}Sr_{0.05}Cu_{0.95}Zn_{0.05}O_4$. The changes of the local spin polarization manifest strongly in the temperature-dependent optical conductivity at ~7.2 eV, with an anomaly at the magnetic stripe phase (~25K), accompanied by anomalous spectral-weight transfer in a broad energy range. Supported by theoretical calculations, we also assign high-energy optical transitions and their corresponding temperature dependence, particularly at ~2.5eV, ~8.7eV, ~9.7eV, ~11.3eV and ~21.8 eV. Our result shows the importance of a strong mixture of spin singlet and triplet states in hole-doped cuprates and demonstrates a new strategy to probe local magnetic correlations using high-energy optical conductivity in correlated electron systems.

## I. Introduction

As a prominent ingredient of the electronic and spin structures and of fundamental relevance for high-temperature superconductivity in copper-oxides (cuprates), an understanding of magnetic and electronic correlations in relation to the charge carriers, i.e. doped holes, in its unusual normal state plays an important role, considering the fact that the superconductivity is induced by carriers doped into the antiferromagnetic insulator [1-6]. After nearly three decades of intensive studies, there is a general consensus, that in the cuprates the doped holes go mainly into O-2$p$ orbitals. This has been experimentally observed by a number of experimental techniques, including X-ray absorption and energy-loss spectroscopies [7-10], angle-resolved photoelectron spectroscopy [11], inelastic neutron scattering [12], and resonance sound velocity measurement [13]. However, how magnetic correlations evolved in the vicinity of doped holes remains a serious issue in the field.

Several theoretical models have suggested the importance of the oxygen sites [2,3,14-16]. However, local descriptions on how the doped hole correlated with the surrounding Cu spins remain hotly-debated among different theoretical models. On one hand, a local singlet character of the doped hole in the $CuO_2$ planes, the so-called Zhang-Rice singlet (ZRS) [2] was proposed. It consists of a doped hole on oxygen and an intrinsic local hole on $Cu^{2+}$ in a singlet wave function with a net zero spin moment. On the other hand, a model of a three-spin polaron described that the doped hole in oxygen can promote local ferromagnetic fluctuations of $Cu^{2+}$ spins surrounding it in an otherwise antiferromagnetic background [3,17]. Therefore, an experimental method that is capable to probe local magnetic correlations in the vicinity of charge carriers is needed to solve this fundamental problem.

The key challenge is to understand the nature of magnetic correlations in lightly hole-doped, high-quality single crystal cuprates even in the proximity of antiferromagnetic regime [18]. Experimentally, because of the small concentration (~5%) of the doped holes, till now it has been very challenging to probe the short-range magnetic correlation around the small number of doped holes at oxygen site, given the overwhelming contribution from the localized Cu spins.

## II. Experimental and Theoretical Methods

Here, we design an experimental approach to address this fundamental problem. A combination of a synchrotron-based experimental technique, i.e. ultraviolet–vacuum ultraviolet (UV-VUV) optical reflectivity and spectroscopic ellipsometry is used to reveal the optical conductivity ($\sigma_1$) of cuprates in an unprecedented energy range up to 32.5eV, with very high accuracy, as a function of temperature and polarization. Note that the high-energy optical conductivity is sensitive to the magnetization as experimentally and theoretically shown in case of manganites [19,20]. This novel technique is applied to high quality *untwinned* single crystal of lightly-doped, non-superconducting $La_{1.95}Sr_{0.05}Cu_{0.95}Zn_{0.05}O_4$ (Zn-LSCO), and undoped $Sr_2CuO_2Cl_2$ (SCOC), as a comparison, which were grown by the traveling solvent floating zone method. The Zn-LSCO has been characterized using neutron scattering revealing that our sample still shows the diagonal incommensurate spin modulation (or diagonal "stripe" phase) below ~25 K ($T_s$) and the role of Zn was found to enhance the magnetic correlations [21]. This physical property is particularly important because; *1)* the antiferromagnetic background is still present in the excess of charge carriers, and *2)* there exists a stripe phase below $T_s$ that demonstrate different local magnetic correlations surrounding the doped hole. The SCOC is isostructural of Zn-LSCO and regarded as a reference system for undoped cuprates [22]. Our analysis and assignments of optical transitions are supported with exact cluster diagonalizations and first-principles density functional calculations.

The normalization procedure of the reflectivity data is the following.

1. *Data collection and simulation*: We collect and combine three set of data: (1) Ψ and Δ from spectroscopic ellipsometry with energy from 0.5 to 6 eV, (2) reflectivity (*R*) from Ultraviolet-vacuum ultraviolet (UV-VUV) reflectance with energy from 3.5 to 35 eV and (3) calculated form factors with energy from 30 eV and well above taken from the Henke table[23]. The first two are from our experiments while the later was from calculated value.

2. *Spectroscopic ellipsometry*: Spectroscopic ellipsometry measures accurately the ellipsometric parameters Ψ (the ratio between the amplitude of *p*- and *s*-polarized reflected light) and Δ (represents the phase difference between *p*- and *s*-polarized reflected light) from which we then calculate reflectivity (R), complex dielectric response and optical conductivity accurately in the energy range of 0.5 to 6 eV without the need of Kramers-Kronig transformation.

3. *The Henke tabulated value*: Rom the Table, one can extract index of refraction (and the atomic form factor) in very broad energy range, from 30 to 30,000 eV. From here, we can calculate reflectivity of a material in that broad energy range.

4. *Normalization*: We then normalize the UV-VUV reflectance spectra to reflectivity from the spectroscopic ellipsometry for low energy side and to calculated reflectivity from the Henke tabulated value for the high energy side. Therefore, we get absolute value of reflectivity from 0.5 to 35 eV. We note that the overlapping in the reflectivity spectrum at the low as well as high energies helps in our normalization procedure.

5. *Fitting*: We then fit the comprehensive reflectivity data using the Lorentz oscillator model that fulfills the Kramers-Kronig transformation. During the fitting procedure, we shall be able to reproduce the results from spectroscopic ellipsometry, e.g. reflectivity, dielectric response, and optical conductivity. This procedure also validates our normalization procedure.

We perform both theoretical calculations, first principles calculations using local spin density approximation and cluster calculations. The theoretical first principles calculations using local spin density approximation (LSDA + U) are performed for $La_2CuO_4$ system, from which, we extract total and partial density of states (DOS) associated with different atomic orbitals, i.e. Cu(1) and Cu(2) refer to two spin antiparallel Cu atoms in the system, O(1) the O atom in the $CuO_2$ plane, O(3) the apical oxygen out of the plane.

The details of cluster calculations are the following. The simplest model to capture the main excitations is a single $CuO_4$ cluster, which includes five 3d orbitals ($d_{x^2-y^2}, d_{z^2-r^2}, d_{xy}, d_{xz}$ and $d_{yz}$) of copper and three 2p orbitals ($p_x, p_y, p_z$) of each oxygen. The undoped case is thus described by the $CuO_4$ cluster with a single hole, and the additional hole doping simply increases the number of holes in the cluster. The cluster model used in this analysis is defined by the following Hamiltonian,

$$H = H_{pot} + H_{t,pd} + H_{t,pp} + H_U$$

(1)

$$H_{pot} = \sum_{s,\alpha} E_\alpha^d d_{\alpha,s}^\dagger d_{\alpha,s} + \sum_{i,\beta,s} E_{i,\beta}^p p_{i,\beta,s}^\dagger p_{i,\beta,s}$$

(2)

$$\begin{aligned} H_{t,pd} &= t_{pd,\sigma} \sum_s (d_{b_{1g},s}^\dagger (-p_{1,x,s} + p_{2,y,s} + p_{3,x,s} - p_{4,y,s}) + \text{h.c.}) \\ &+ t_{pd,z} \sum_s (d_{a_{1g},s}^\dagger (-p_{1,x,s} - p_{2,y,s} + p_{3,x,s} + p_{4,y,s}) + \text{h.c.}) \\ &+ t_{pd,\pi} \sum_s (d_{b_{2g},s}^\dagger (p_{1,y,s} + p_{2,x,s} - p_{3,y,s} - p_{4,x,s}) + \text{h.c.}) \\ &+ t_{pd,\pi} \sum_s (d_{e_{gx},s}^\dagger (p_{1,z,s} - p_{3,z,s}) + \text{h.c.}) \\ &+ t_{pd,\pi} \sum_s (d_{e_{gy},s}^\dagger (p_{2,z,s} - p_{4,z,s}) + \text{h.c.}) \end{aligned}$$

(3)

$$\begin{aligned} H_{t,pp} &= \frac{t_{pp,\sigma}}{2} \sum_{i=1}^{4} \sum_s ((p_{i+1,x,s}^\dagger + (-1)^i p_{i+1,y,s}^\dagger)(p_{i,x,s} + (-1)^i p_{i,y,s}) + \text{h.c.}) \\ &- \frac{t_{pp,\pi}}{2} \sum_{i=1}^{4} \sum_s ((p_{i+1,x,s}^\dagger + (-1)^{i+1} p_{i+1,y,s}^\dagger)(p_{i,x,s}^\dagger + (-1)^{i+1} p_{i,y,s}^\dagger) + \text{h.c.}) \end{aligned}$$

(4)

$$\begin{aligned} H_U &= \frac{U}{2} \sum_{\alpha,s} n_{d,\alpha,s} n_{d,\alpha,\bar s} + \frac{U'-J}{2} \sum_{\alpha \neq \alpha',s} n_{d,\alpha,s} n_{d,\alpha',s} + \frac{U'}{2} \sum_{\alpha \neq \alpha',s} n_{d,\alpha,s} n_{d,\alpha',\bar s} \\ &- \frac{J}{2} \sum_{\alpha \neq \alpha',s} d_{\alpha,s}^\dagger d_{\alpha,\bar s} d_{\alpha',\bar s}^\dagger d_{\alpha',s} + \frac{J}{2} \sum_{\alpha \neq \alpha',s} d_{\alpha,s}^\dagger d_{\alpha,\bar s}^\dagger d_{\alpha',\bar s} d_{\alpha',s} \\ &+ \frac{U_p}{2} \sum_{i,\beta,\beta',s,s'} p_{i,\beta,s}^\dagger p_{i,\beta,s} p_{i,\beta',s'}^\dagger p_{i,\beta',s'} + V_{pd} \sum_i \sum_{\alpha,\beta,s,s'} d_{\alpha,s}^\dagger d_{\alpha,s} p_{i,\beta,s}^\dagger p_{i,\beta,s} \end{aligned}$$

(5)

where $d^\dagger_{\alpha,s}$ ($d_{\alpha,s}$) creates (annihilates) a hole with orbital $\alpha$ ($\alpha = b_{1g}, a_{1g}, b_{2g}, e_{gx}, e_{gy}$) and spin ($s = \pm\frac{1}{2}$) on copper, and $n_{d,\alpha,s} = d^\dagger_{\alpha,s} d_{\alpha,s}$. Note that $b_{1g}, a_{1g}, b_{2g}, e_{gx}$, and $e_{gy}$ correspond to $d_{x^2-y^2}, d_{3z^2-r^2}, d_{xy}, d_{xz}$ and $d_{yz}$ orbitals, respectively, which form the bases of the irreducible representations of point group symmetry $D_{4h}$. $p^\dagger_{i,\alpha,s}$ ($p_{i,\alpha,s}$) is the creation (annihilation) operator of a hole at the $i$-th oxygen (see Figure S4) with $\alpha$ (= $x, y, z$) orbital and spin $s$. Here, we impose that $p^\dagger_{5,\alpha,s} = p^\dagger_{1,\alpha,s}$. $H_{pot}$ is the on-site potential energy term with $E^p_{1,y} = E^p_{2,x} = E^p_{3,y} = E^p_{4,x} = E^p_{p\pi}$, and $E^p_{i,z} = E^p_{pz}$, due to the crystal-field effect. $H_{t,pd}$ and $H_{t,pp}$ are the hopping terms for the nearest neighbor Cu-O bonds and the nearest neighbor O-O bonds, respectively. $H_U$ is the interaction term including the on-site intra-orbital Coulomb repulsion on the copper $U$, the on-site inter-orbital Coulomb repulsion $U'$, the Hund's Coupling $J$, the pair hopping term $J'$ at the Cu site, the on-site Coulomb interaction $U_p$ at O site, and the nearest neighbor Coulomb interaction $V_{pd}$ between Cu and O sites. The parameters used in this study are shown in **Table 1**. Note that these parameters are matched with our experimental results and are consistent to those reported by previous papers [15,24].

### III. Result and Discussion

**In Fig. 1**, we show high-energy reflectivity [**Figs. 1(a)** and **1(b)**] and high-energy optical conductivity, σ₁ [**Figs. 1(c)** and **1(d)**] of Zn-LSCO as a function of temperature (from 8 to 300 K) and polarization [**E**∥*a*\* and **E**∥*b*\*, see inset of **Fig. 1(a)**], together with SCOC for **E**⊥*c* as reference. Intriguingly, new optical transitions are observed at high-energies (>3eV), i.e. a pronounced peak at 7.2eV and a rather weak peak at 21.8eV. Supported by theoretical calculations, these new peaks originate from excitations involving doped holes in the oxygen orbitals of the Cu-O planes; they only occur in Zn-LSCO and they are absent in the undoped reference sample even though the overall spectrum of the optical conductivity is very similar showing that it is dominated by the copper-oxygen planes. Furthermore, we also observe other high-energy optical transitions, e.g. at ~8.7, ~9.7, and ~11.3eV, which occur in both samples but with some smaller differences in their details. At low energies (< 3eV), a well-known charge transfer gap is observed in both, hole doped Zn-LSCO at ~2.5eV and undoped SCOC at a slightly lower energy of ~2.0eV. However, only in the hole-doped Zn-LSCO we observe a mid-infrared response (~0.7eV), resulting from the holes doped into the system [25].

The key observation of our measurement is the surprisingly strong temperature dependence of the 7.2eV peak as clearly shown in the change of optical conductivity, $\Delta\sigma_1$ [**Figs. 1(e)** and **1(f)**]. We observe a spectral-weight transfer as high as 15% [**Figs. 1(g)** and **1(h)**] with an anomaly around $T_s$. As the temperature decreases, these peaks steadily enhance. Intriguingly, when the diagonal stripe phase develops at $T_s$ [21], an abrupt change occurs and these peaks decrease dramatically as temperature further decreases. Such a new optical anomaly near the stripe phase is important as a model system to reveal the local spin configurations surrounding the doped holes. Correspondingly, in accordance to the first-moment sum rule, a large temperature-dependent spectral-weight loss is observed in the energy range of 7.6 - 20.9eV, which is three order magnitude higher than any thermal energy scale ( <~30meV). This strongly points towards strong electronic correlations. As an example the bare Coulomb on-site repulsion in cuprates is of the order of 10-20 eV [26]. The total integrated spectral weight is conserved to within 0.2%, which allows us to reveal magnetic and charge correlations in cuprates.

We perform exact cluster diagonalization and first-principles density functional calculations within local spin density plus U (LSDA+U) approximation. We are mainly interested in the identification of the basic optical transitions which are mostly the intercluster excitations. These are basically represented by the electron removal states of one cluster combined with the electron addition states of a neighboring cluster, which we consider to be uncoupled from each. Let us first calculate the one-electron removal and one-electron addition spectral functions $A_\alpha(\omega) = A_\alpha^+ + A_\alpha^-$ for the undoped case. Here

$$A_\alpha^+(\omega) = -\frac{1}{\pi} \lim_{\delta \to 0^+} \text{Im} \langle \psi_0 | c_\alpha^\dagger \frac{1}{\omega - H + E_0 + i\delta} c_\alpha | \psi_0 \rangle \tag{6}$$

and

$$A_\alpha^-(\omega) = -\frac{1}{\pi} \lim_{\delta \to 0^+} \text{Im} \langle \psi_0 | c_\alpha \frac{1}{\omega - E_0 + H + i\delta} c_\alpha^\dagger | \psi_0 \rangle, \tag{7}$$

where $c_\alpha (= d_\alpha, p_\beta)$ corresponds to the annihilation operator of hole, and $|\psi_0\rangle$ is the ground state of the $CuO_4$ cluster with a single hole (see **Fig. 2**). As shown in **Fig. 3**, all excitations are classified by using their symmetry. For instance, one particle electron-removal spectral function, corresponding to the electron removal of a $x^2$-$y^2$ symmetry electron resulting in two holes in $CuO_4$, with $A_{1g}$ symmetry are shown in **Fig. 3(c)**.

Theoretical first principles calculations using local spin density approximation (LSDA + U) were performed for $La_2CuO_4$ system, from which, we extract total and partial density of states (DOS) associated with different atomic orbitals, i.e. Cu(1) and Cu(2) refer to two spin antiparallel Cu atoms in the system, O(1) the O atom in the CuO$_2$ plane, O(3) the apical oxygen out of the plane, as shown in **Fig. 4**.

The resulting energies of different electronic states and their spin polarization for the local Copper and Oxygen sites are shown in **Figs. 5(a)** and **5(b)**, for an undoped cuprate. Form this it is clear that spin-polarization is favored for O(1) states suggesting a strong triplet contribution [c.f. **Fig. 5(a)**]. Consistent with cluster calculations as shown in **Fig. 5(b)**, the main transition comes from the Cu-$d_{x^2-y^2}$, which is ~7.2 eV below the Fermi Level. Based on the optical transition rule, the hole doping is then essential to activate this optical transition (see discussion below). Therefore, we would not expect to see this state in undoped but rather in slightly doped cuprates. Pictorial many-body schemes are shown in **Figs. 2(c)** and **(d)**. **Figure 2(c)** describes a Cu-$d^9$ site with spin pointing up (one spin-down hole, as shown in the plot). In this configuration, the ZRS appears below the chemical potential as it is possible to create a ZRS by adding a spin-up hole to the oxygen site. **Figure 5(d)** describes a Cu-$d^9$ with a ligand hole in the ZRS configuration, so it is possible to destroy the ZRS by adding an electron (above the chemical potential) in either spin channel.

The high-energy optical conductivity involves high-energy states of the Cu-$d^8$ orbitals, which are well-known to exist in the 7 to 15eV range (see also Refs. [27,28] for other cuprates), and can be understood via inter-site transitions involving O and the neighboring two Cu sites. The most relevant microscopic processes of the transition at 7.2eV, showing a strong temperature dependence, are illustrated in **Fig. 6**. They are dominated by transitions involving O-2$p$ orbitals and the σ-bonding Cu-3$d_{x^2-y^2}$ or Cu-

$3d_{3Z^2-r^2}$ orbitals. Since one of the holes on Cu-$d^8$ must be a $d_{x^2-y^2}$ hole, we consider only the $d^8$ states with two $d_{x^2-y^2}$ holes, which form spin singlets or one $d_{x^2-y^2}$ and one $d_{3Z^2-r^2}$ hole, which can also form a triplet. In a pure singlet scenario, the 7.2 eV peak should be independent from temperature because in such scenario the peak originated largely from excitations of a O-2*p* doped hole centered on one Cu (in a ZRS) to a neighboring Cu-3*d* orbital forming a singlet state $d^8(S)$: | $d^9$ ZRS > → | $d^8(S) d^9$> [see **Fig. 6b**]. While other transitions, i.e. ~8.7, ~9.7, and ~11.3eV and 21.8eV features, which are present at both Zn-LSCO and SCOC samples, are dominated by the Copper-Oxygen planes as discussed below.

The strong temperature dependence of 7.2eV peak reveals an important and yet surprising nature of the ZR state, namely the Cu and O orbitals being actually spin-polarized rather than spin-neutral as in a pure singlet form [see also **Fig. 2(a)**]. Note that in strongly correlated materials, even spin-preserved charge response of electrons is still constrained by the spin correlation of the system due to the Pauli exclusion principle. Hence the wave function has to contain now a mixed singlet and triplet (MST) wave function [**Figs. 2(e)** and **3(c)**], instead of the pure singlet [**Figs. 2(d)** and **3(b)**]. Since the local ferromagnetic correlations are expected to be enhanced as the temperature is being reduced till $T_s$, the temperature evolution of the 7.2eV peak is a direct example of this scenario [**Fig. 3(c)**]. Only the component of the wavefunction that would have ferromagnetically aligned Cu spins, can contribute to the temperature-dependent optical transition matrix element [c.f. **Figs. 3(c)** and **3(d)**]. Therefore, an increase of the spectral-weight of this excitation with decreasing temperature indicates an increase of a ferromagnetic correlation between the surrounding Cu spins, reflecting the influence of the O-2$p$ hole.

This somewhat unexpected conclusion can be understood from the consideration of the correlation between the MST and the neighboring Cu spins [**Fig. 5(c)**]. In the MST, the symmetry in the spin degree of freedom is broken because the surrounding Cu spins are strongly polarized in one direction. Consequently, the MST benefits from a ferromagnetically aligned component between neighboring Cu sites. This allows maximizing the virtual kinetic processes [c.f. purple arrows **Fig. 5(c)**] between the doped hole at O sites and the intrinsic hole in the neighboring Cu sites. Hence, the process leads effectively to a mixing of a triplet component into the wave function. This is the same microscopic process that leads to the formation of the three-spin polaron model in the O-centered local picture [16,17]. Obviously, the more the surrounding Cu spins align, the stronger this effect is, and the stronger the 7.2eV peak grows [c.f. inset of **Figs. 1(c)** and **1(d)**] showing that the ferromagnetic alignment of the neighboring Cu spins seems to be the lowest energy state since the ferromagnetic correlations increase with decreasing temperatures.

Our measurement addresses an important issue on the low-temperature magnetic structure at $T < T_s$. Interestingly, the new 7.2eV peak is also very sensitive to the formation of a long-range ordered "stripe" phase [21], i.e. the abrupt reversal of the trend in the spectral weights near $T_s$, which can be explained within MST picture. The stripe correlation hosts an anti-phase boundary of the anti-ferromagnetic correlation across the doped hole [29-34]. That is, the Cu atoms on the opposite side of the MST are correlated with opposite spin. This leads to a compensation of the net magnetic moment of the surrounding spins of the MST and consequently to the observed abrupt decrease of intensity in the 7.2eV transition. Thus, the stripe correlation starts to reduce the ferromagnetic correlation across the doped hole of MST at $T < T_s$.

More generally, the observed triplet component of the MST due to the polarization of surrounding spins implies an important change of a minimum model for the cuprates. For example, upon integrating out the triplet states, one typically arrives at the so-called *t-J* model by dropping many "non-essential" high-energy terms in the process. Our observation indicates that higher order processes like $-\left(\sum_{neighbor} \vec{S}\right)^2 h^+ h$ or $-\sum_{neighbor} \vec{S} \cdot h^+ \vec{\sigma} h$ would need to be included, with $h^+$ ($h$) denoting creation (annihilation) of doped holes and $\vec{S}$ the spin of the surrounding Cu. Interestingly, the corresponding physical effect of such terms is a tendency to align Cu spin ferromagnetically near doped holes *without* moving the doped hole, an effect emphasized by Emery and Reiter [17]. Previously, such a tendency for the development of ferromagnetic correlations is derived only from the kinetic motion of the doped holes, instead of an effective potential.

In the next discussions, we describe the optical transitions and their corresponding temperature dependent at ~2.5 eV, ~8.7eV, ~9.7eV, ~11.3eV and ~21.8eV (see **Fig. 7**). We start our discussion on the basic electronic band structure for undoped cuprate. It is generally accepted that the parent compound SCOC, which is similar to $La_2CuO_4$, is an antiferromagnetic insulator with spin of 1/2 on Cu and with a charge-transfer type conductivity gap of ~2eV (see **Fig. 1**). This fixes the energy scales of the first electron addition and removal states, consistent within the pictorial model shown in **Fig. 6(a)** as discussed later. The band width of the electron addition state is roughly 1eV as determined from LSDA+U calculations (see **Fig. 4**) and the dispersion width of the first electron removal states determined from angular resolved photoelectron spectroscopy is about 0.3eV or roughly twice the superexchange interaction between the local Cu spins [35]. This narrow electron removal structure is referred to as the Zhang-Rice singlet (ZRS), which is composed of one hole in a Cu-$d_{x^2-y^2}$ orbital and one hole in a linear combination of bonding O-2$p$ orbitals also with $x^2 - y^2$ symmetry around the central Cu [**Fig. 3(b)**]. At higher electron removal energies the remaining O-2$p$ orbitals form bands, which are ~5eV wide, based on LSDA+ U calculations covering an energy range to ~5eV below the Fermi level that is fixed at the top of the valence band (see also **Fig. 4**). A similar feature is also obtained in the cluster calculations [see **Fig. 3(b)**]. Interestingly, the LSDA+U calculations show that O-2$p$ is spin-polarized. At even higher electron removal energies ranging from ~3.5 to ~12eV below the Fermi level, spin-resolved photoemission and Auger spectroscopies as well as satellite structures in photoemission spectroscopy [27,28] have been found and identified as Cu-$d^8$ states, which are spread over an energy range ~8eV due to the atomic multiplet structure resulting from the large atomic Coulomb and exchange-like interactions between the two $d$ holes (see also **Fig. 3**).

  The 8.7 eV feature is optical excitations involving a hole from a ZRS to a neighboring Cu which ends up in a $d^8$ ($d_{x^2-y^2}$ and $d_{3z^2-r^2}$) spin triplet state (**Fig. 8**). If we consider transitions into $d^8$ triplet states, this transition is expected to show a quite different temperature behavior, i.e. $\sigma_1$ increases as temperature decreases. Transitions into $d^8$ triplet states are considerably weaker in intensity involving $d_{3z^2-r^2}$ orbitals, but more importantly the corresponding transition to a $d^8$ triplet state from the doped hole plaquette that has been influenced by the transition at 9.7eV which is involving a $d^8$ singlet to the UHB and has an opposite temperature dependence in the $\sigma_1$. If, however, we could observe the transition to the $d^8$ triplet state we would be using the same arguments as above but expect it to have the reversed temperature dependence.

The 9.7 eV feature involves optical excitations of a hole from a central Cu in a $d^9$ ($d_{x^2-y^2}$) state to a neighboring Cu which ends up in a $d^8$ spin singlet state (**Fig. 9**). The temperature dependence of $\sigma_1$ involving the $d^8$ singlet and the Upper Hubbard band would obviously display an increase of $\sigma_1$ with decreasing temperature because here the low-temperature state would surely involve a strong antiferromagnetic alignment of the neighboring Cu spins. On the other hand, the ferromagnetic alignment has no contribution to this transition due to the Pauli principle.

The 11.3 eV feature originates mainly from transitions of a hole from a central Cu in $d^9$ without the presence of a ZRS to a neighboring Cu $d^8$ spin triplet state involving the UHB (**Fig. 10**). Here, we observe rather strong temperature dependence, i.e. the $\sigma_1$ decreases as temperature decreases. This requires a starting state with the Cu spins parallel for the largest $\sigma_1$. However the ground state is clearly one where these spins are antiparallel. The change of the spectral weight transfer of the 11.3eV feature for $T \leq T_s$ is related to the stripe formation to compensate for the change of the spectral weigh transfer of the polarized mixed singlet and triplet (MST) (see also main article). The main point is that optical transitions involving the $d^8$ states can result in either triplet or singlet local states, which is fundamentally different from what can happen in the single band Hubbard model.

The 21.8 eV feature corresponds to excitation of a doped hole from ZRS to the neighboring O-2$s$ core level: $|d^9\text{ ZRS}> \rightarrow |s^1 d^9 d^9>$ [**Fig. 11(a)**]. With the presence of an effective exchange coupling between the O-2$s$ level with the O-2$p$ unpaired spin and indirectly with the Cu spin, the feature ~21.8eV, which shows a reduction at the lower energy side around 19.6eV and an enhancement at the higher energy side around 21.8eV seen in $\Delta\sigma_1$ [c.f. **Figs. 1(e)** and **1(f)**], can be understood using the proposed MST model as illustrated in **Fig. 11**. Our model well explains that the optical transition from O-2$s$ to the local spin-polarization results in a different sign in the change of the total spectral weight around 21.8eV [**Figs. 11(c)** and **11(d)**]. While, the unpolarized ZRS would be insensitive to changes of local magnetic correlations [**Fig. 11(b)**].

## IV. Conclusion

In conclusion, we reveal a strong mixture singlet and triplet configuration in the lightly-hole doped La$_{1.95}$Sr$_{0.05}$Cu$_{0.95}$Zn$_{0.05}$O$_4$ single crystal using high-energy optical conductivity. The doped hole is shown to induce the effective MST wave function that enhances ferromagnetic correlations between Cu spins near the doped holes. Our result also demonstrates a new strategy and potency of high-energy optical conductivity to locally probe the interplay of magnetic correlations surrounding the doped holes in strongly correlated electron systems.

## V. Acknowledgment

We acknowledge Anthony J. Leggett and G. Baskaran for valuable discussions. This work is supported by Singapore National Research Foundation under its Competitive Research Funding (NRF-CRP 8-2011-06), MOE-AcRF Tier-2 (MOE2015-T2-1-099), NUS-YIA, FRCs, BMBF through VUVFAST (05K2014) as well as DFG through Ru 773/4-1 and RIKEN. W.K. acknowledges support from National Natural Science Fundation of China (NSFC) #11447601, and Ministry of Science and Technology (MOST) #2016YFA0300500 and 2016YFA0300501. The works at University of British Columbia are supported by NSERC, QMI, and CIfAR. The works at work at Tohoku University are supported by Grants-in-Aid from The Ministry of Education, Culture, Sports, Science and Technology (MEXT), Japan (No. 23340093).

**FIGURE AND TABLE CAPTIONS**

**FIG. 1. (a-b)** Reflectivity and **(c-d)** optical conductivity, $\sigma_1$ of lightly hole-doped La$_{1.95}$Sr$_{0.05}$Cu$_{0.95}$Zn$_{0.05}$O$_4$ (Zn-LSCO) as a function of temperature and light polarizations as indicated in the figures. The undoped Sr$_2$CuO$_2$Cl$_2$ (SCOC) is used for comparison. The vertical-dashed lines show the new excitations at 7.2eV. The inset of **(a)** shows CuO-plane with CuO$_4$ plaquettes in the orthorhombic $a^*$ and $b^*$ axes. The inset of **(c)** and **(d)** is a magnification of excitations near 7.2eV. (e-f) The change of optical conductivity $\Delta\sigma_1(T)$ and (g-h) relative changes of integrated spectral-weight (SW). The $\Delta\sigma_1(T)$ is defined as $\Delta\sigma_1(T) - \Delta\sigma_1(T=300K)$. The relative change of the integrated spectral-weight $\frac{SW(T)}{SW(300K)}$ is defined as $\frac{\int_{\omega_1}^{\omega_2}\Delta\sigma_1(\omega,T)\,d\omega}{\int_{\omega_1}^{\omega_2}\Delta\sigma_1(\omega,T=300K)\,d\omega}$, where $T$ is temperature and $\omega_1$ ($\omega_2$) is the photon energy at $\omega_1$ ($\omega_2$). The integrated spectral weight for two different regions – SW$_{II}$ (($\omega_1$ to $\omega_2$) = (from 5.0eV to 7.8eV)) or as SW$_{7.2eV}$, and SW$_{total}$ (from 0.5eV to 32.5eV) are shown. The critical temperature for the diagonal stripe order is indicated by $T_s$. The overall spectral weight $SW(T)$ from 0.5 to 32.5eV is conserved within 0.2%.

**FIG. 2.** Schematic figure of a CuO$_4$ cluster. Black and white circles indicate copper and oxygen, respectively.

**FIG. 3.** Projected one-electron removal and one-electron addition spectral functions $A_\alpha(\omega)$ for the undoped CuO$_4$ cluster. The symmetry $\Gamma_\alpha$ (in $D_{4h}$) for orbital α is indicated in the figures **(c)-(f)**. Since the ground state of the undoped system is of $b_{1g}$ symmetry, the symmetry of the one electron removal (addition) states is easily obtained by $\Gamma_\alpha \times b_{1g}$. The total symmetries for the electron removal states of **(c)** A$_{1g}$ and **(e)** B$_{1g}$ represent the two hole states of current interest. Orange and blue lines indicate triplet and singlet states, respectively. For comparison, the one-electron removal spectra for all **(a)** Cu 3$d$ and **(b)** O 2$p$ orbitals are also shown. The states below zero (the Fermi energy indicated by solid lines) are the electron removal state and those above zero are the electron addition states. The ZR singlet (UHB) is indicated by black (red) dashed lines.

**FIG. 4.** Theoretical first principles calculations using LSDA + U for parent $La_2CuO_4$ system. **(a)** Total and **(b)-(f)** partial density of states (DOS) associated with different atomic orbitals. Cu(1) and Cu(2) refer to two spin antiparallel Cu atoms in the system, O(1) the O atom in the $CuO_2$ plane, O(3) the apical oxygen out of the plane. The energy range of the plot are **(a)-(d)** -32.5 to 17.5 eV and **(e)-(f)** -8 to 5 eV.

**FIG. 5. (a)** Density of states based on local spin density approximation (LSDA) + U for Cu-$d_{x^2-y^2}$ and O-$p$ of parent compound $La_2CuO_4$ system. **(b)** Projected one-electron removal and one-electron addition spectral functions $A_\alpha(\omega)$ for the undoped $CuO_4$ cluster and the total symmetries for the electron removal states of $A_{1g}$. The symmetry $\Gamma_\alpha$ (in $D_{4h}$) for orbital α is indicated. Since the ground state of the undoped system is of $b_{1g}$ symmetry, the symmetry of the one electron removal (addition) states is obtained by $\Gamma_\alpha \times b_{1g}$. The total symmetries for the electron removal states of $A_{1g}$ represents the two hole states of current interest [36]. **(c)** Conceptual local density of states (LDOS) based on experimental result and calculations with local spin density approximations for the undoped case. Acronyms S, T and ZRS stand for Singlet, Triplet and Zhang-Rice Singlet states, respectively. $d^n$ ($p^n$) indicates the number $n$ of $d$ ($p$) electrons in Cu (O). The multiple level in the Cu-$d^8$ ($d^8(S)p^6$) singlet configuration is due to the crystal field splitting. (We note that the same symbols are used in Figs. 3 and 4). (Left) Spin structure, (middle) LDOS and (right) local wave function for **(d)** unpolarized, stand-alone ZRS and **(e)** strongly spin polarized of mixed singlet triplet state (MST). LDOS (↓) denotes local density of state for spin-down. Note that the same symbols are used in **Fig. 3**.

**FIG. 6. (a)** Schematic hole configurations of the 7.2eV optical transition within two $CuO_4$ plaquettes. The left (right) two $CuO_4$ plaquettes represent the initial (final) state of the corresponding optical excitation. The $d^9$ plaquette state corresponds to the ground state of the $CuO_4$ plaquette with a single hole and the ZRS plaquette corresponds to a state with two holes, one in Cu and one in O sites. The energy of the optical excitation is calculated from the $CuO_4$ plaquette model, matched with experimental data. **(b)** An inter-site optical transition and excitation spectrum of unpolarised ZRS corresponding to the $CuO_4$ plaquettes configuration in **(a)**. The optical transition occurs by exciting an electron

from (*top panel*) $d^8(S)p^6$ state of $d^9(\downarrow)$ CuO$_4$ plaquette into (*middle panel*) the unpolarized ZRS CuO$_4$ plaquette. (*Bottom panel*) This optical transition is allowed only for spin up configuration, therefore the resulting excitation spectrum is for spin up configuration. Such an optical transition does not change as a function of temperature because it does not depend on surrounding spin configurations. **(c)** An inter-site optical transition and excitation spectrum of polarized mixed singlet and triplet state (MST). The optical transition occurs by exciting an electron from (*top panel*) $d^8(S)p^6$ state of $d^9(\downarrow)$ CuO$_4$ plaquette into (*middle panel*) the polarized MST CuO$_4$ plaquette. (*Bottom panel*) This optical transition is allowed only for spin up configuration with a strong temperature dependence. As temperature decreases, such an optical excitation increases because the local ferromagnetic correlation enhances, consistent with experimental data. **(d)** The temperature dependence in the MST configuration yields a robust temperature dependent optical transition at 7.2eV as temperature decreases, fully consistent with our experimental result (c.f. **Fig. 1**). LDOS ($\downarrow$) denotes local density of state for spin-down of hole.

**FIG. 7.** Spectral weight analysis for different incoming light polarizations, **E**||($a^*,b^*$), and different spectral regions of La$_{1.95}$Sr$_{0.05}$Cu$_{0.95}$Zn$_{0.05}$O$_4$. **(a)** and **(b)** Change of optical conductivity $\Delta\sigma_1(T)$ defined as $\sigma_1(T) - \sigma_1(T = 300K)$ and **(c)** and **(d)** relative change of the integrated spectral weight $\frac{SW(T)}{SW(300K)}$ defined as $\frac{\int_{\omega_1}^{\omega_2} \Delta\sigma_1(\omega,T)\, d\omega}{\int_{\omega_1}^{\omega_2} \Delta\sigma_1(\omega,T=300K)\, d\omega}$, where $T$ is temperature (in Kelvin) and $\omega_1$ ($\omega_2$) is the photon energy at $\omega_1$ ($\omega_2$). We show the integrated spectral weight for seven different regions – SW$_\text{I}$ (($\omega_1$ to $\omega_2$) = (from 0.5eV to 5.0eV)), SW$_\text{II}$ (from 5.0eV to 7.8eV) or as SW$_\text{7.2ev}$ in the main text, SW$_\text{III}$ (from 7.8 eV to 9.2eV), SW$_\text{IV}$ (from 9.2 eV to 18.0eV), SW$_\text{V}$ (from 18.0eV to 20.8eV), SW$_\text{VI}$ (from 20.8eV to 21.5eV) or as SW$_\text{21.8ev}$ in the main text, and SW$_\text{total}$ (from 0.5eV to 32.5eV)**.** $\Delta\sigma_1(T)$ and $SW(T)/SW(300K)$ for different polarization and spectral regions are indicated in the figures**.** The critical temperature for the diagonal stripe order is indicated by $T_{s[21]}$. In the supplementary, we focus our discussion on SW$_\text{III}$ for explaining the σ$_1$ at 8.7eV and on SW$_\text{IV}$ for 9.7eV and 11.3eV. The overall spectral weight $SW(T)$ from 0.5 to 32.5eV is conserved within 0.2%.

**FIG. 8.** Pictorial model of the electronic band structure and optical transitions at 8.7eV. **(a)** Schematic electronic and spin configurations of CuO$_4$ plaquettes consisting of a doped hole forming a Zhang-Rice singlet (ZRS). In each set of figures, the left (right) two CuO$_4$ plaquettes represent the initial (final) state of the corresponding optical excitation. The $d^9$ plaquette state corresponds to the ground state of the CuO$_4$ plaquette with a single hole and the ZRS plaquette corresponds to a state with two holes, one in Cu and one in O sites. Assuming each plaquette is independent, the optical excitation energy is easily estimated from the CuO$_4$ plaquette model. Excitation spectrum for **(b)** transition to unpolarized ZRS and **(c)** transition to polarized mixed singlet and triplet state (MST) [c.f. **Figs. 6 (b)** and **(c)**]. LDOS ($\downarrow$) and LDOS ($\uparrow$) and denote local density of state for spin-down and spin-up, respectively. **(d)** Change of total spectral weight as a function of temperatures expected from this model. Note that the more complete pictorial model of electronic band structure can be seen at **Fig. 5(a).**

**FIG. 9.** Pictorial model of the electronic band structure and optical transitions at 9.7eV. **(a)** Schematic electronic and spin configurations of CuO$_4$ plaquettes. In each set of figures, the left (right) two CuO$_4$ plaquettes represent the initial (final) state of the corresponding optical excitation. The $d^9$ plaquette state corresponds to the ground state of the CuO$_4$ plaquette with a single hole. Assuming each plaquette is independent, the optical excitation energy is easily estimated from the CuO$_4$ plaquette model. The $S$ denotes singlet configuration. **(b)** Excitation spectrum for transition to Upper Hubbard band in antiferromagnetic correlation. LDOS ($\downarrow$) and LDOS ($\uparrow$) and denote local density of state for spin-down and spin-up, respectively. **(c)** Change of total spectral weight as a function of temperatures expected from this model. Note that the more complete pictorial model of electronic band structure can be seen in **Fig. 5 (a)**.

**FIG. 10.** Pictorial model of the electronic band structure and optical transitions at 11.3eV. Schematic electronic and spin configurations within a CuO$_4$ plaquette for **(a)** transition to upper Hubbard band with ferromagnetic correlations which occurs at $T_s$ and **(b)** transition to Upper Hubbard band in antiferromagnetic correlation which occurs at low

temperatures. In each set of figures, the left (right) two CuO₄ plaquettes represent the initial (final) state of the corresponding optical excitation. The $d^9$ plaquette state corresponds to the ground state of the CuO₄ plaquette with a single hole. Assuming each plaquette is independent, the optical excitation energy is easily estimated from the CuO₄ plaquette model. The *T* and *S* denote triplet and singlet configuration, respectively. Excitation spectrum for **(c)** transition in the phase with ferromagnetic correlations and **(d)** transitions in the phase with antiferromagnetic correlation. LDOS (↓) and LDOS (↑) and denote local density of state for spin-down and spin-up, respectively. **(e)** Change of total spectral weight as a function of temperatures expected from this model. Note that the more complete pictorial model of electronic band structure can be seen in **Fig. 5(a).**

**FIG. 11.** Pictorial model of the electronic band structure and optical transitions at 21.8eV. **(a)** Schematic hole (↓) configurations of the 21.8eV optical transition within two CuO₄ plaquettes involving semi-core O-2*s* orbital. The calculated optical excitation energy is matched with experimental data. Note that the explanation of symbols and notations is referred to **Figs. 2** and **3**. **(b)** An inter-site optical transition from (*top panel*) semicore O-2*s* to (*middle panel*) unpolarised ZRS and (*bottom panel*) corresponding excitation spectra for both spin up and spin down configurations based on the CuO₄ plaquettes configuration in **(a)**. This optical transition is allowed for both spin up and spin down configurations, the resulting excitation spectrum has equal spectral-weight in the unpolarized ZRS scheme. As a result, this transition does not change as a function of temperature because it does not depend on surrounding spin configurations. **(c)** An inter-site optical transition from (*top panel*) semicore O-*s* to (*middle panel*) spin-polarized mixed singlet and triplet states (MST) and (*bottom panel*) corresponding excitation spectra for both spin up and spin down configuration. As temperature decreases, the local ferromagnetic configuration enhances. This yields to an increased spectral-weight for spin up and a decreased spectral-weight for spin down. **(d)** Based on MST configurations, an increased spectral-weight at ~21.8eV side is accompanied with a decreased spectral-weight at ~20eV, fully consistent with our experimental results (c.f. **Figs. 1(e)** and **1(f)**). Note that the O-2*s* orbital corresponds to the $d_{x^2-y^2}$ symmetric superposition of four O-2*s* orbitals surrounding the Cu and is therefore orthonormal between each site. It splits

because of strong Hunds coupling and $|d^8 L>$ orbital, with $L$ as a ligand hole. The Cu-$d$ and O $p$ are both more occupied in the spin up channel, thus lowering the energy of O 2$s$ in the spin up channel.

**Table 1.** The parameter set used is listed. Here (pd$\sigma$) [(pd$\pi$)] is Slater-Koster parameter for $\sigma(\pi)$ bonding between $p$ and $d$ orbitals [36]. These paramaters are taken to match our experimental data.

# FIGURES

**FIG. 1**

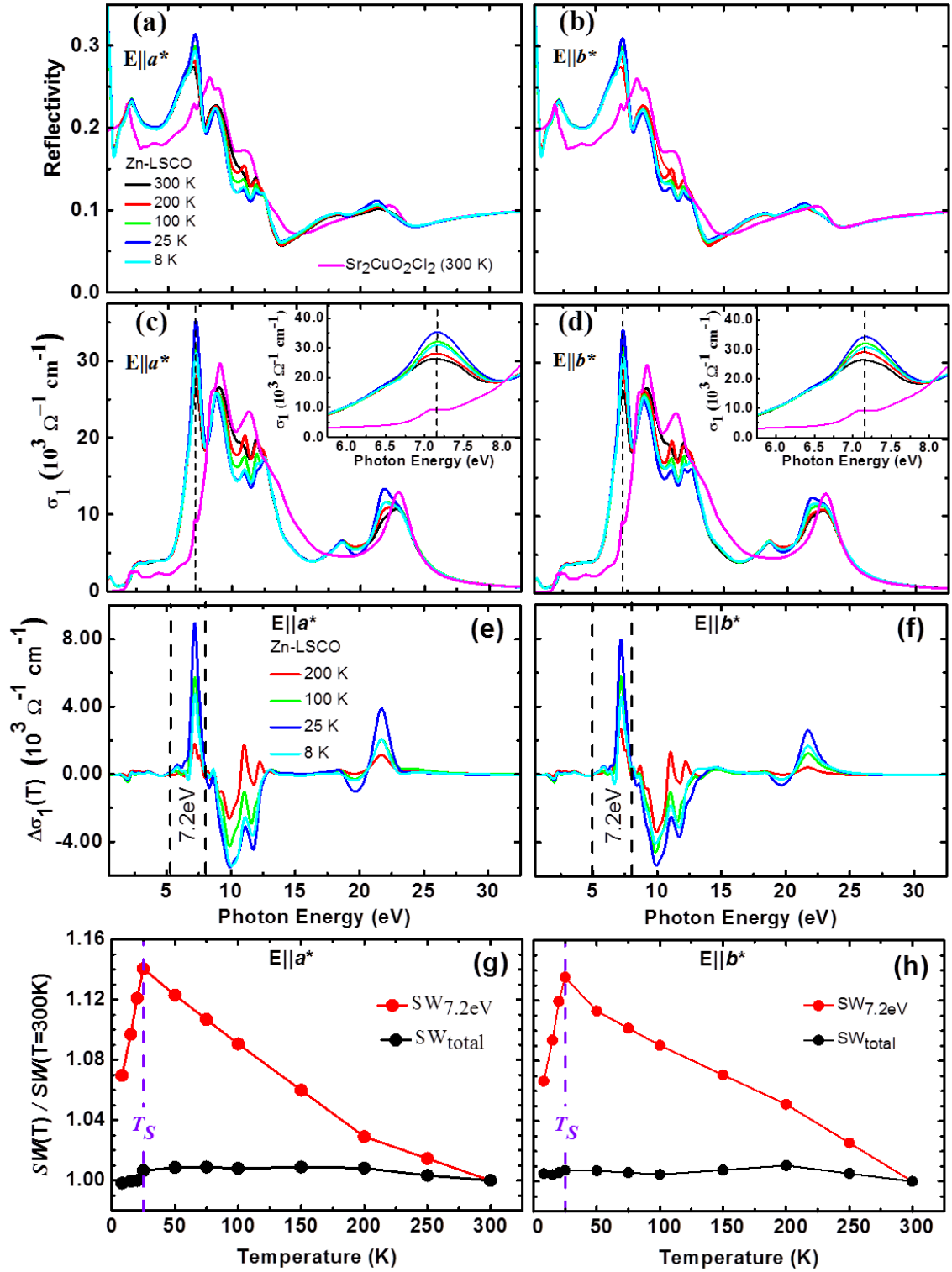

**FIG. 2**

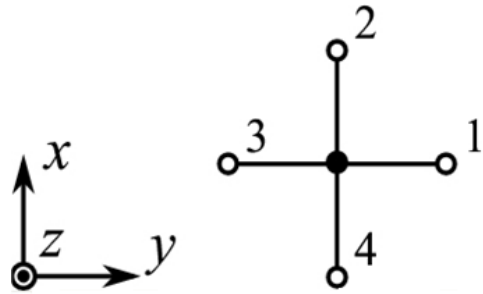

**FIG. 3**

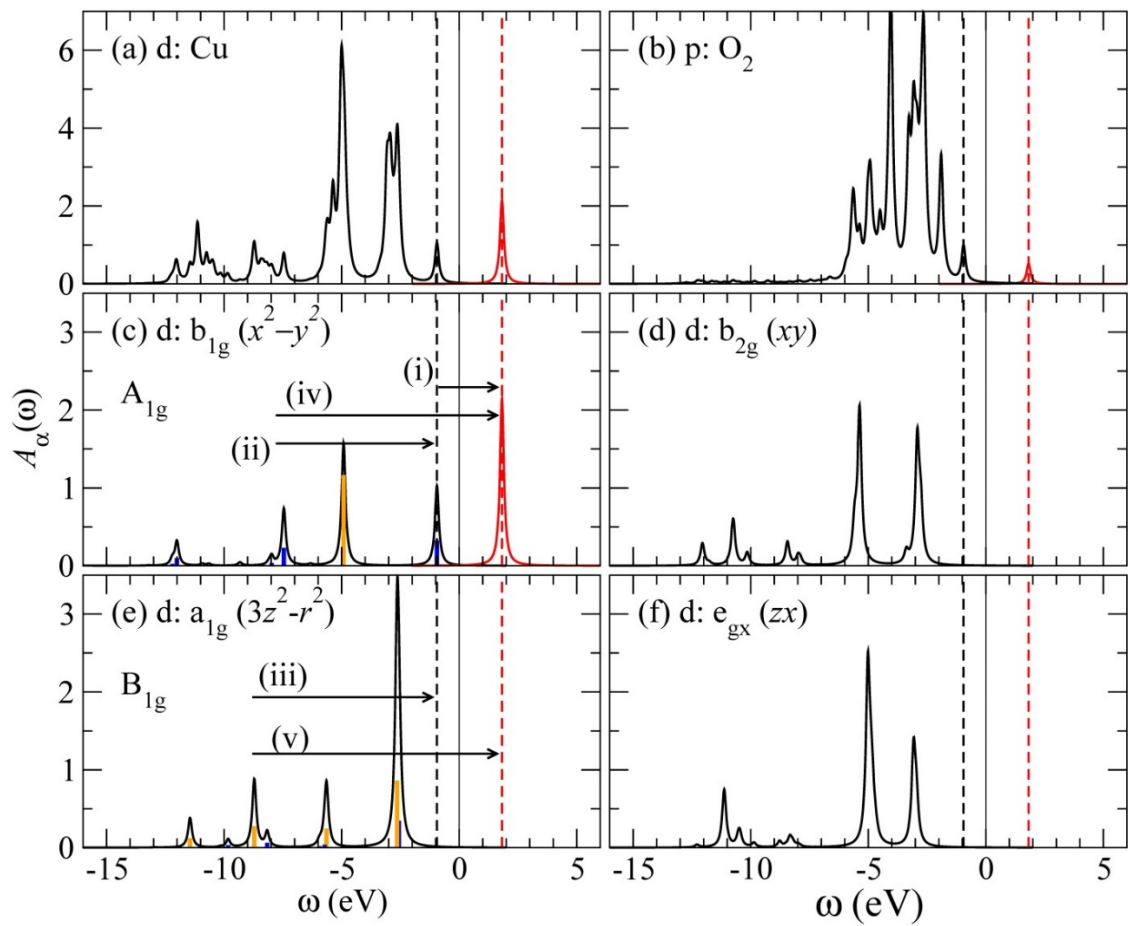

**FIG. 4**

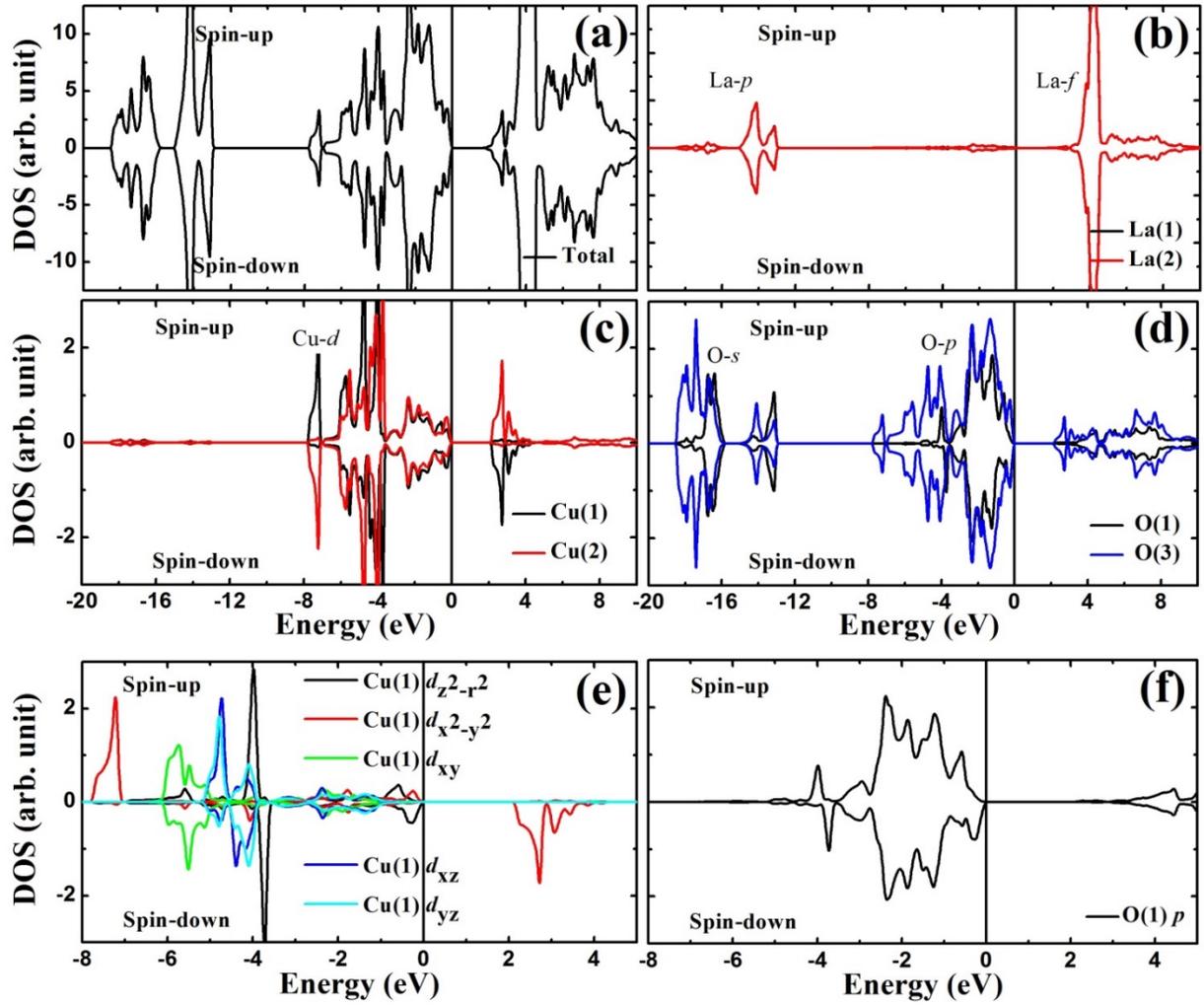

**FIG. 5**

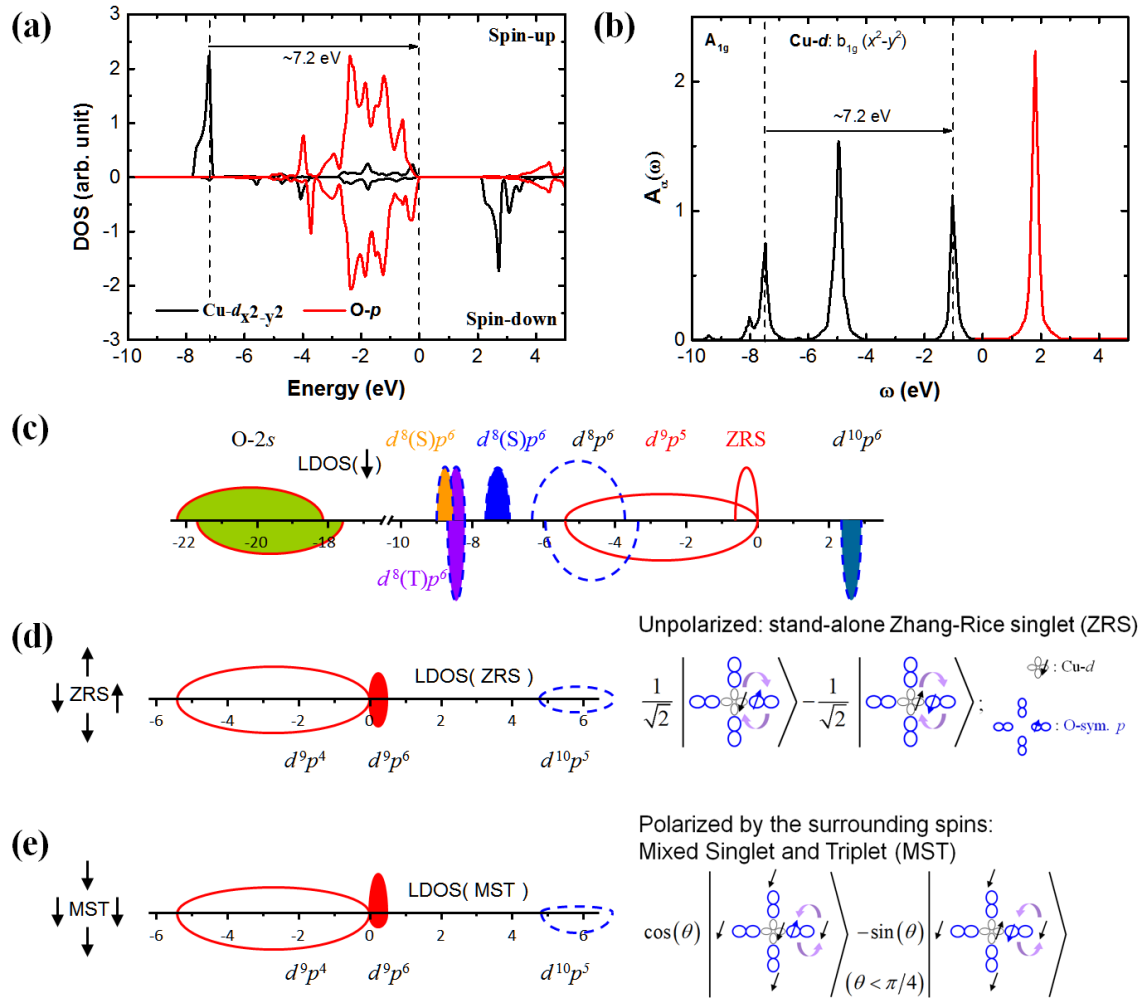

**FIG. 6**

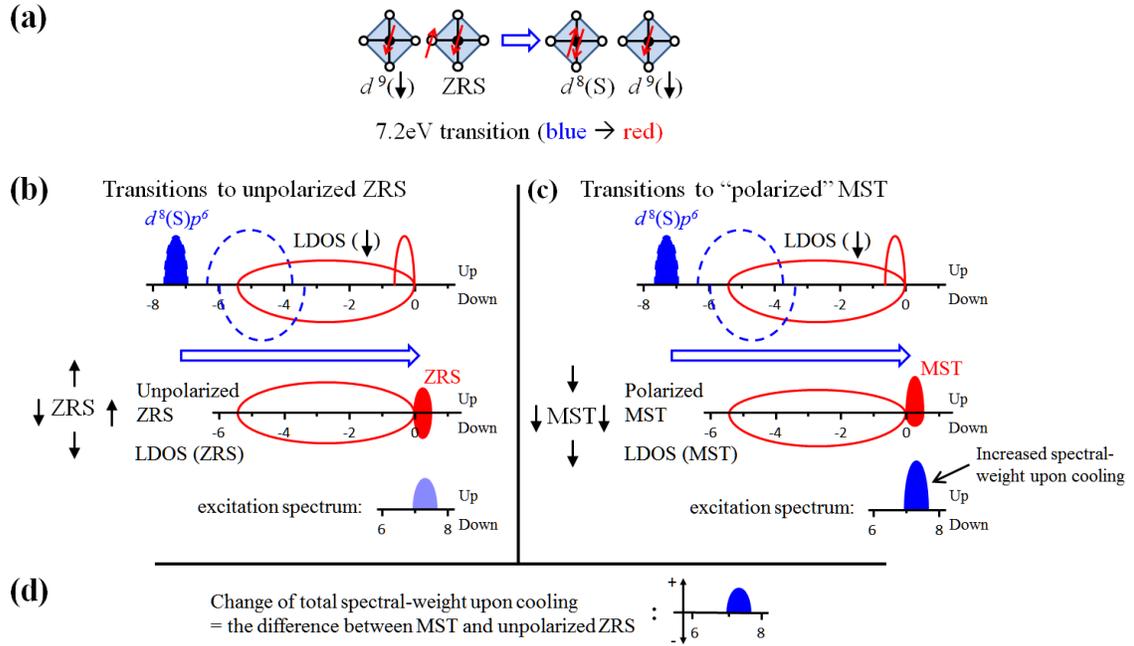

**FIG. 7**

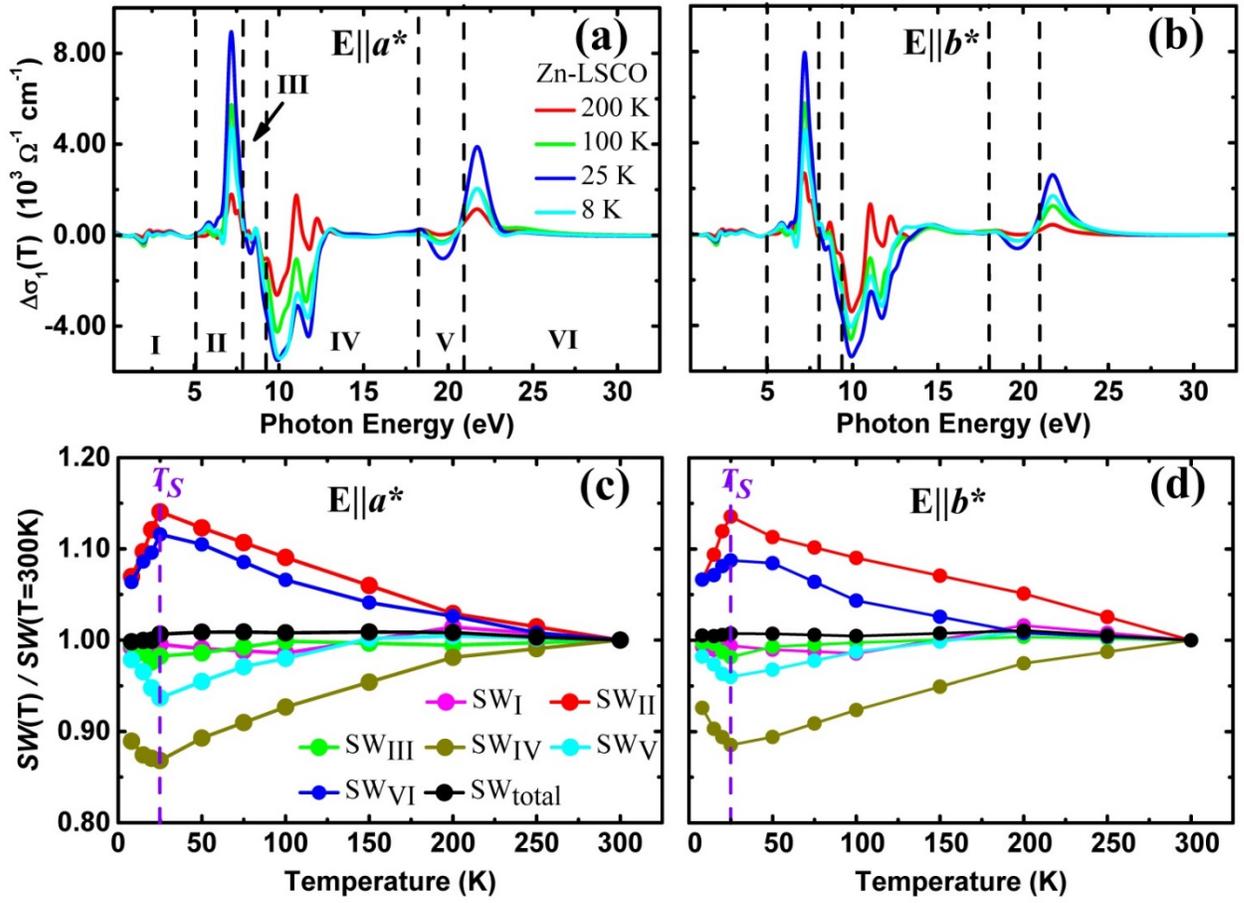

**FIG. 8**

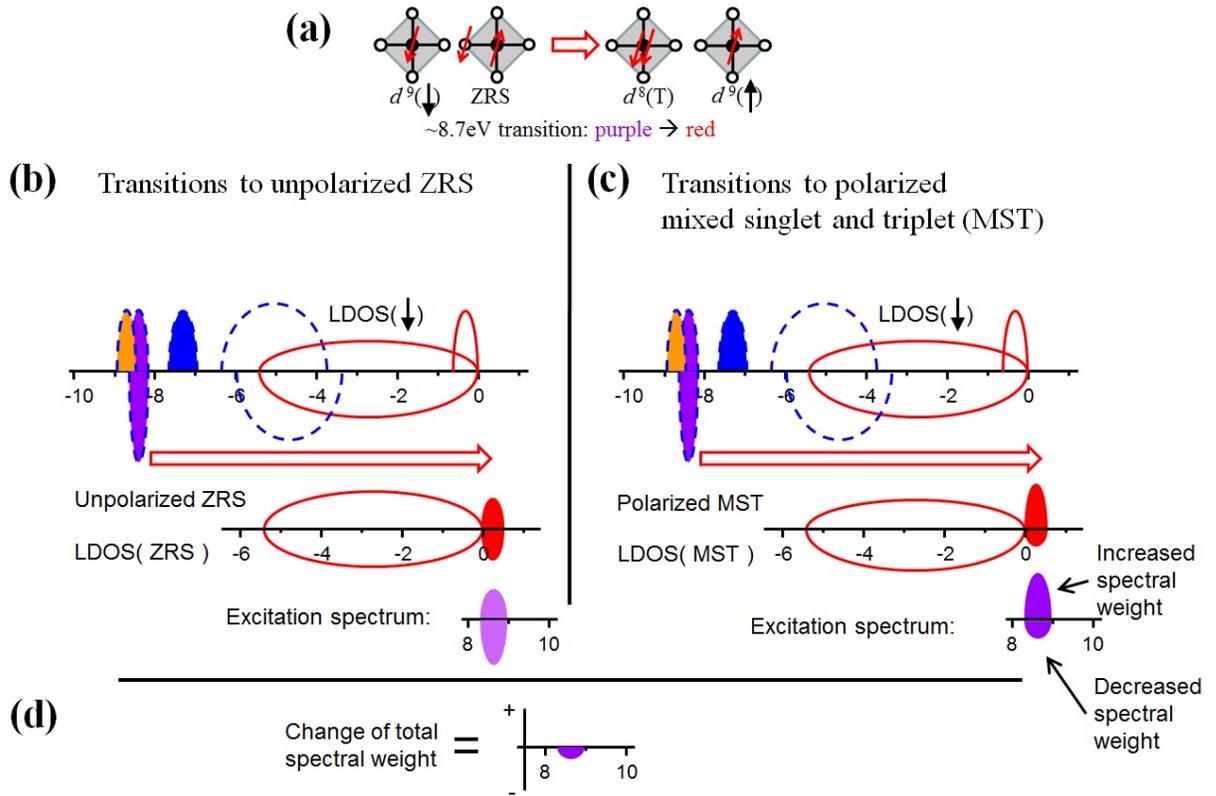

**FIG. 9**

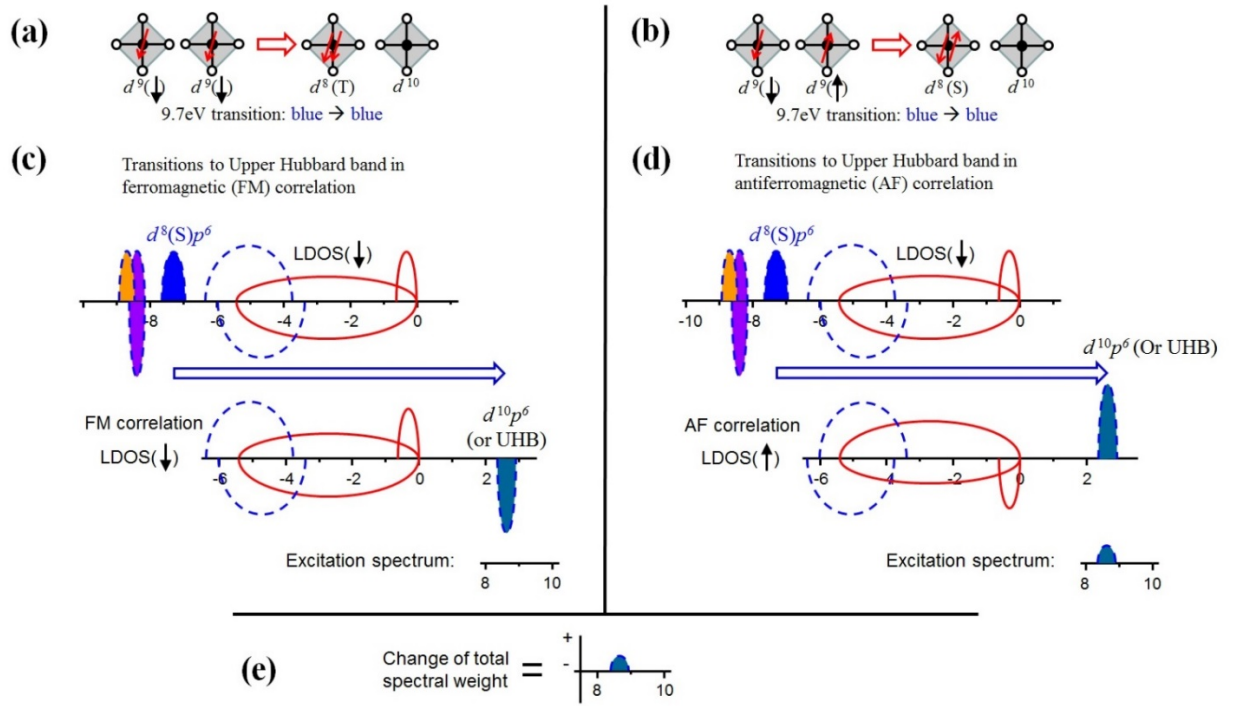

**FIG. 10**

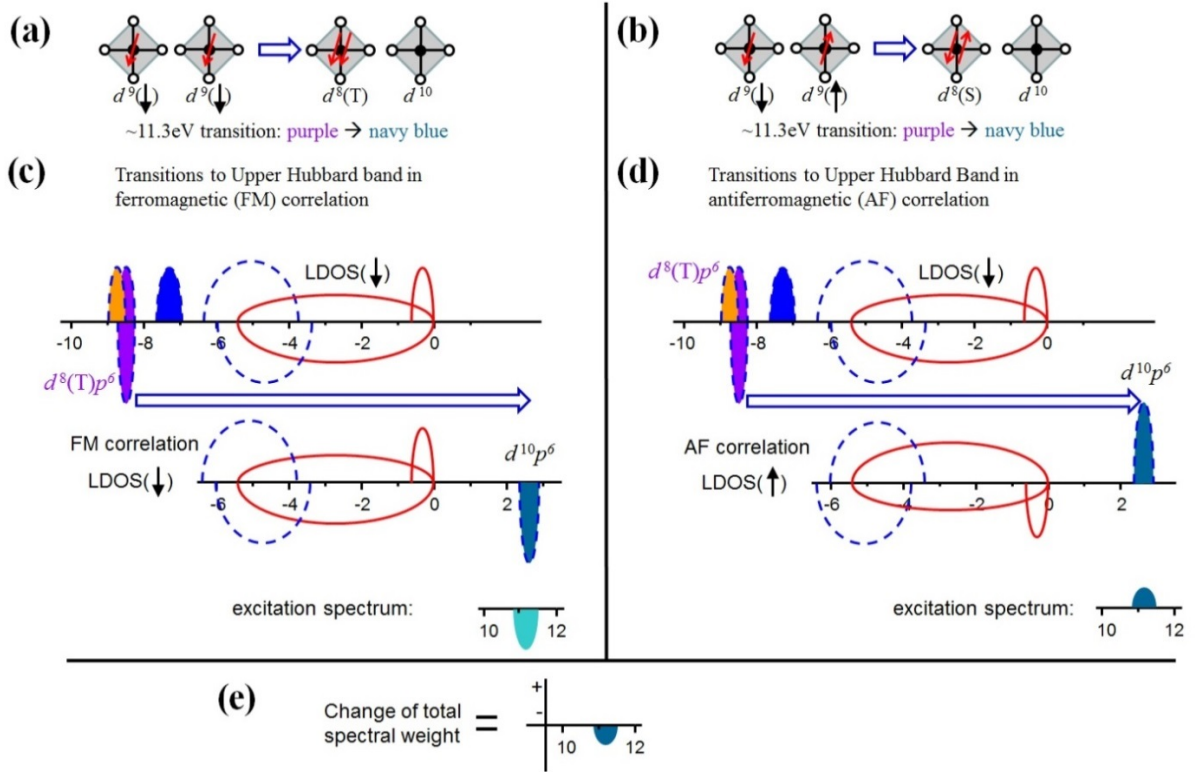

**FIG. 11**

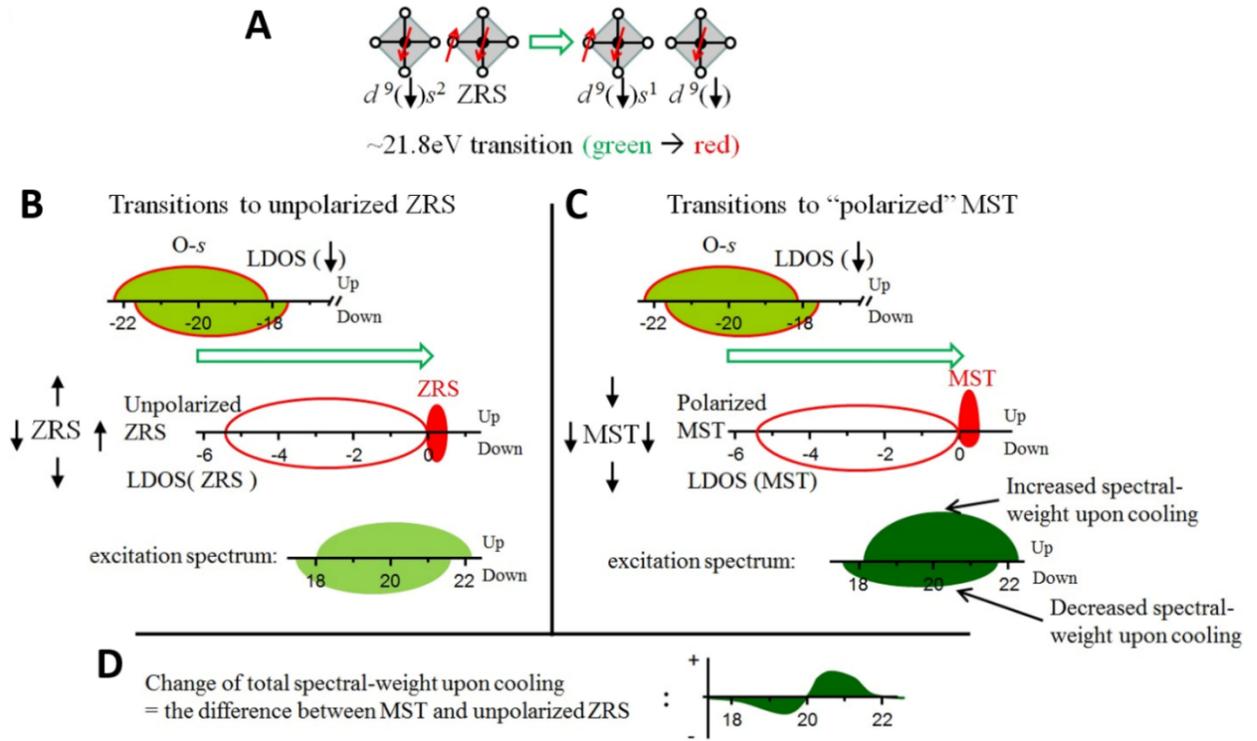

**TABLE I.**

| Parameter | Value (eV) |
|---|---|
| $t_{pd\sigma}(=\sqrt{3}(pd\sigma)/2)$ | 1.3 |
| $t_{pd\pi}(=(pd\pi))$ | -0.7 |
| $t_{pdz}(=(pd\sigma)/2)$ | 0.65 |
| $t_{pp\sigma}$ | -1.0 |
| $t_{pp\pi}$ | 0.3 |
| $E_{b_{1g}}, E_{a_{1g}}$ | 0 |
| $E_{b_{2g}}, E_{e_g}$ | 0.9 |
| $E_{p\sigma}$ | 3.0 |

| Parameter | Value (eV) |
|---|---|
| $E_{p\pi}$ | 2.0 |
| $E_{pz}$ | 2.7 |
| $U$ | 8.8 |
| $U'$ | 6.5 |
| $J$ | 1.2 |
| $J'$ | 1.2 |
| $U_p$ | 4.0 |
| $V_{pd}$ | 1.2 |